\documentstyle[11pt,amssymb]{article}

\makeatletter
\@addtoreset{equation}{section}
\makeatother

\def\dalemb#1#2{{\vbox{\hrule height .#2pt
        \hbox{\vrule width.#2pt height#1pt \kern#1pt
                \vrule width.#2pt}
        \hrule height.#2pt}}}

\def\cao{{\cal O}}

\def\0{{\sst{(0)}}}
\def\1{{\sst{(1)}}}
\def\2{{\sst{(2)}}}
\def\3{{\sst{(3)}}}
\def\4{{\sst{(4)}}}
\def\5{{\sst{(5)}}}
\def\6{{\sst{(6)}}}
\def\7{{\sst{(7)}}}
\def\8{{\sst{(8)}}}

\def\ep{\epsilon}
\def\td{\tilde}

\def\half{{\textstyle{1\over2}}}

\let\a=\alpha   \let\d=\delta

 \let\t=\tau    
  \let\D=\Delta  
    
    \let\G=\Gamma
\let\la=\label  
  
\def\nn{\nonumber} \def\bd{\begin{document}} \def\ed{\end{document}}
\def\ds{\documentstyle} \let\fr=\frac \let\bl=\bigl \let\br=\bigr
\let\Br=\Bigr \let\Bl=\Bigl
\let\bm=\bibitem
\let\na=\nabla
\let\pa=\partial \let\ov=\overline
\newcommand{\be}{\begin{equation}}
\newcommand{\ee}{\end{equation}}
\def\ba{\begin{array}}
\def\ea{\end{array}}
\def\ft#1#2{{\textstyle{{\scriptstyle #1}\over {\scriptstyle #2}}}}
\def\fft#1#2{{#1 \over #2}}
\def\del{\partial}
\def\sst#1{{\scriptscriptstyle #1}}
 \def\oneone{\rlap 1\mkern4mu{\rm l}}
\def\ie{{\it i.e.\ }}
\def\via{{\it via}}
\def\semi{{\ltimes}}
\def\str{{\rm str}}
\def\Dm{{{D_{\sst{max}}}}}
\def\vac{ \left | 0 \right \rangle }
\def\kvac{ \left | k \right \rangle }

\def\sp{\; \; \;}

\def\bol{ \left | B (p^+) \right \rangle}
\def\bo1{ \left | B^0 (p^+) \right \rangle}

\def\bolt{ \left | B (p^+) \right \rangle_{\t}}

\def\boxl{ \left | B (x^-) \right \rangle}

\def\<{ \langle }
\def\>{ \rangle }

\def\S{\Sigma}

\newcommand{\hsp}{\hspace{0.5cm}}

\newcommand{\ho}[1]{$\, ^{#1}$}
\newcommand{\hoch}[1]{$\, ^{#1}$}
\newcommand{\bea}{\begin{eqnarray}}
\newcommand{\eea}{\end{eqnarray}}
\newcommand{\ra}{\rightarrow}
\newcommand{\lra}{\longrightarrow}
\newcommand{\Lra}{\Leftrightarrow}
\newcommand{\ap}{\alpha^\prime}
\newcommand{\bp}{\tilde \beta^\prime}
\newcommand{\tr}{{\rm tr} }
\newcommand{\Tr}{{\rm Tr} }
\newcommand{\NP}{Nucl. Phys. }

\newcommand{\ams}{{\it Institute for Theoretical Physics,
University of Amsterdam, \\
Valckenierstraat 65, 1018XE Amsterdam, The Netherlands} \\
{\tt taylor@science.uva.nl}}

\newcommand{\auth}{\large Marika Taylor}

\begin{document}
\begin{flushright}
\hfill{ITFA-2007-35}
\end{flushright}

\vspace{15pt}

\begin{center}

{\Large \bf Matching of correlators in AdS$_3$/CFT$_2$}

\vspace{20pt}

\auth

\vspace{15pt}

\vspace{8pt}

{\ams}

\vspace{15pt}

\underline{ABSTRACT}
\end{center}

Recently exact agreement has been found between three-point
correlators of (single particle) chiral operators computed in string theory on $AdS_3
\times S^3 \times T^4$ with NS-NS flux and those computed in the
symmetric orbifold CFT. However, it has also been shown that these
correlators disagree with those computed in supergravity, under any
identification of single particle operators which respects the symmetries. In
this note we resolve this disagreement: the key point is that mixings
with multi-particle operators are not suppressed even at large $N$ in extremal correlators.
Allowing for such mixings, orbifold/string theory operators and supergravity
operators can be matched such that both non-extremal and extremal three point functions agree,
giving further evidence for the non-renormalization of the chiral ring.

\noindent

\pagebreak

\section{Introduction and summary}

Chiral primary operators play an important role in testing
the AdS/CFT correspondence. Supersymmetric chiral primaries
have protected dimensions, and matching between CFT spectra at
weak coupling and supergravity spectra at strong coupling provided the
earliest checks of AdS/CFT.

In the correspondence between ${\cal N} = 4$ SYM and string theory on
$AdS_5 \times S^5$, the three point functions of $1/2$ BPS (single
trace) operators computed from supergravity were found to match the
corresponding correlators computed in free field theory \cite{Lee:1998bx}.
This indicated the existence of a previously unknown non-renormalization
theorem for such correlators which was subsequently proved, modulo
various subtleties, in \cite{Heslop:2001gp}. Moreover, although four and higher point
functions of chiral primaries are in general renormalized, there is
evidence that extremal correlators, in which the dimension of one
operator is equal to the sum of the others, are also protected
\cite{D'Hoker:1999ea}; see also the review \cite{D'Hoker:2002aw}.

It is natural to ask whether similar properties for correlators of
chiral primaries hold in the case
of AdS$_3$/CFT$_2$ dualities. The simplest such case is the D1-D5
system, with $n_1$ D1-branes and $n_5$ D5-branes. Here
the duality is between type IIB in an $AdS_3 \times S^3
\times X_4$ background, where $X_4$ is either $T^4$ or $K3$, and a
two-dimensional ${\cal N} = 4$ superconformal field theory; see for
example the review \cite{David:2002wn}.

The bulk and boundary theories in this case are known to have
equivalent moduli spaces \cite{Dijkgraaf:1998gf,Larsen:1999uk},
but they are tractable only at distinct points in the moduli space.
In the bulk one can work in the supergravity limit, as one does in the case of
$AdS_5 \times S^5$. One can also consider
the S-dual system without RR flux, where the string theory is
tractable: for Euclidean $AdS_3$
it is described by $H_3^{+}$ and $SU(2)$ WZW models at
level $k = n_5$. The boundary theory becomes tractable in the orbifold
limit, namely when the SCFT becomes the symmetric orbifold theory
with target space $N = n_1 n_5$ copies of $X_4$. Note that the
orbifold theory is not the boundary theory
dual to the weakly curved, weakly coupled RR $AdS_3 \times S^3
\times X_4$ background; the boundary theory is a marginal deformation
of the orbifold theory, in which the orbifold is resolved.

Whilst the limits in which the boundary and bulk theories are
tractable are at different points in the moduli space, matching of
the spectrum of chiral primaries is still possible.
Comparison of the spectra obtained from
supergravity with those of the boundary theory was first carried out
in \cite{de Boer:1998ip, Deger:1998nm}. There were also early
attempts to compare three point functions computed from supergravity
with those computed in the orbifold CFT. Extremal three point
functions were computed in the orbifold CFT in
\cite{Jevicki:1998bm,Lunin:2001pw} whilst the cubic couplings in
supergravity relevant for computing three point functions were
determined in \cite{Mihailescu:1999cj, Arutyunov:2000by, Pank}. It
was however noted that these cubic couplings do not match in
structure the extremal three point functions computed in the
orbifold theory. Only the cubic couplings in supergravity, and not
the three point functions, were computed in \cite{Arutyunov:2000by,
Pank}. Computing the three point functions is rather subtle, in that
systematic holographic renormalization \cite{S2} is required to
obtain the correct correlators, satisfying the requisite Ward
identities.

Moreover, extremal correlators
are subject to additional subtleties: the bulk extremal cubic couplings
vanish, and the corresponding three point functions are obtained from
finite boundary terms in the action, which in turn should follow from careful
reduction of the ten-dimensional action \cite{D'Hoker:1999ea}. Put
differently, one should first include boundary terms in the
ten-dimensional action such that the variational problem is well-posed
for the appropriate Dirichlet boundary conditions, and then dimensionally
reduce to obtain the effective three-dimensional action.

In practice, it is more convenient to obtain the
extremal correlators by analytic continuation of the corresponding
non-extremal correlators. That is, one defines the extremal three
point functions as
\bea
\< {\cal O}_{\D_2 + \D_3} (x_1) {\cal O}_{\D_2} (x_2) {\cal O}_{\D_3}
(x_3) \> = \frac{C_{(\D_2 + \D_3) \D_2 \D_3}}{ | \vec{x_1} -
  \vec{x_2}|^{2 \D_2} | \vec{x_1} -
  \vec{x_3}|^{2 \D_3}}; \nn \\
C_{(\D_2 + \D_3) \D_2 \D_3} = {Lim}_{\Delta_1 \rightarrow (\D_2
    + \D_3)} ( C_{\D_1 \D_2 \D_3}),
\eea
where the scalar operator ${\cal O}_{\D}$ has dimension $\D$, and the
non-extremal structure constant $C_{\D_1 \D_2 \D_3}$ follows from the
bulk non-extremal couplings. This analytic continuation was discussed
in \cite{D'Hoker:1999ea} and more recently such a definition of
extremal correlators was discussed in \cite{Skenderis:2006uy} within the framework
of holographic renormalization. Note that this approach implicitly
assumes that the structure constants are analytic in the operator
dimensions, which need not be true, given that the latter are
discrete.

Holographically renormalized non-extremal correlators for
scalar chiral primaries were recently computed from supergravity in
\cite{Kanitscheider:2006zf}, and the corresponding extremal
correlators were then determined via analytic continuation.
These extremal correlators were compared to those computed in the orbifold
CFT. Since only a subset of non-extremal correlators of scalar chiral
primaries have so far been computed in the orbifold theory, in
\cite{Lunin:2001pw}, only extremal correlators could be compared.

A structural disagreement between these correlators was found. To be more precise,
the single particle scalar chiral primaries in the orbifold CFT are labeled by the $(p,p)$
cohomology of $X_4$, their twist $n \ge 1$, and their R symmetry
$SU(2)_L \times
SU(2)_R$ quantum numbers as
\be \la{o-1}
{\cal O}^{0,0}_{n m \bar{m}}; \qquad
{\cal O}^{(r)1,1}_{n m \bar{m}}; \qquad
{\cal O}^{2,2}_{n m \bar{m}}.
\ee
Here $(m,\bar{m})$ are the eigenvalues of $J^3$ and $\bar{J}^3$
respectively and $(r)$ labels the $(1,1)$ cohomology of $X_4$, of
dimension $h^{1,1}$; thus $(r)$
runs from $1$ to $4$ for $T^4$ and from $1$ to $20$ for $K3$. The
operator dimension $\Delta$ is related to the twist and cohomology via
\be
\Delta = (n - 1 + p),
\ee
with $J = \bar{J} = \half \Delta$ being the $SO(4)$ R-symmetry quantum numbers.
The cohomology label implicitly defines the transformation
properties under the $SO(h^{1,1})$ global symmetry of the CFT.

On the supergravity side one has a set of operators dual to scalar fields
in $AdS_3$ which are labeled by their dimension $\Delta$ and R symmetry $SU(2)_L \times
SU(2)_R$ quantum numbers:
\be \la{o-2}
{\cal O}^{S^{(a)}}_{\Delta m \bar{m}}; \qquad
{\cal O}^{\Sigma}_{\Delta m \bar{m}},
\ee
where $\Delta \ge 1$ for ${\cal O}^{S^{(a)}}_{\Delta m \bar{m}}$ and
$\Delta \ge 2$ for ${\cal O}^{\Sigma}_{\Delta m \bar{m}}$. Here $(a)$
runs from $0$ to $h^{1,1}$ and $\Phi \equiv (S^{(a)}, \S)$ are the bulk
scalar fields, which couple to these operators. Of the $(h^{1,1} + 1)$
operators ${\cal O}^{S^{(a)}}_{\Delta m \bar{m}}$ one transforms as a
singlet under the $SO(h^{1,1})$ global symmetry and the remaining
$h^{1,1}$ as a vector. $ {\cal O}^{\Sigma}_{\Delta m \bar{m}}$ is also
a singlet under the $SO(h^{1,1})$ symmetry; see \cite{Deger:1998nm} for further
details, and tables of the operators.

Already from (\ref{o-1}) and (\ref{o-2}) one can see a subtlety in
comparing correlators: the identification between orbifold CFT
operators and those dual to supergravity fields is not unique, since
the protected quantum numbers of dimension, R symmetry charge and the $SO(h^{1,1})$
global symmetry leave some degeneracy. Given that ${\cal O}^{\S}_{1 m \bar{m}}$ is
absent, a natural identification between orthonormal operators is
\bea
{\cal O}^{0,0}_{n m \bar{m}} & \leftrightarrow & {\cal O}^{S}_{(n-1) m
  \bar{m}}; \\
{\cal O}^{(r)1,1}_{n m \bar{m}} & \leftrightarrow & {\cal O}^{S^{(r)}}_{n m
  \bar{m}}; \nn \\
{\cal O}^{2,2}_{n m \bar{m}}  & \leftrightarrow & {\cal O}^{\S}_{(n+1) m
  \bar{m}}, \nn
\eea
and it is this identification which has been assumed in previous literature,
but any linear rotation of this identification such that
\be
\left ( \begin{array} {c} {\cal O}^{S}_{(n-1) m   \bar{m}} \\
{\cal O}^{\S}_{(n-1) m   \bar{m}} \end{array} \right )
= {\cal M}
\left ( \begin{array} {c} {\cal O}^{0,0}_{n m \bar{m}} \\
{\cal O}^{2,2}_{(n-2) m \bar{m}} \end{array} \right ),
\ee
with ${\cal M}$ an arbitrary $SO(2)$ matrix also respects the
symmetries and orthonormality.

The result of \cite{Kanitscheider:2006zf} was that there is a disagreement in
the extremal correlators for any choice of ${\cal M}$. The
disagreement is structural in that for any such linear identification
many more of the correlators are non-vanishing in the orbifold theory
than in supergravity. Such a disagreement is a priori perhaps not surprising:
the computations are at different points in the moduli space and there
is no known non-renormalization theorem. Even in the ${\cal N} = 4$ SYM
theory, which has thirty-two supercharges, the non-renormalization
of the analogous three point functions is rather subtle and the proof requires
assuming the absence of certain conformal invariants
\cite{Heslop:2001gp}. Renormalization in this case, with only sixteen
supercharges, is not a priori excluded, particularly as the orbifold
theory is a marginal deformation of the actual boundary CFT.

However, recently three point functions of chiral primaries were
computed in the NS-NS $AdS_3 \times S^3 \times X_4$ background using
the WZW model description of the worldsheet theory.
In \cite{Dabholkar:2007ey} extremal three point functions of all
single particle chiral primaries were computed, whilst in \cite{Gaberdiel:2007vu}
non-extremal three point functions for operators in the
${\cal O}^{0,0}_{n m \bar{m}}$ family were computed. Later in
\cite{Pakman:2007hn} the
calculations were extended to non-extremal three point functions of
all chiral primaries. All of these correlators agree exactly with those
computed in the orbifold CFT, although let us recall that in the latter
only a subset of the non-extremal correlators have so far been determined.

That computations at different points in the moduli space agree indicates that
there is indeed a non-renormalization theorem protecting these
correlators, but at the same time raises the puzzle of why the
extremal correlators computed from supergravity did not agree with the
orbifold CFT (and hence the string) computations. Whilst it is undoubtedly
interesting to find that there is a non-renormalization theorem
protecting these correlators, it is arguably more important to understand whether there
is any unresolved subtle issue in comparing supergravity and dual field theory
results. The reason is that in many situations in gravity/gauge theory
dualities one wants to use the supergravity description
as a tool to compute the strong coupling result, exactly, when no
non-renormalization theorem applies.

In this paper we will resolve this issue, and explain how the
supergravity correlators are reconciled with the orbifold CFT and string
theory correlators. The conclusions are the following. All {\it non-extremal} three
point functions computed via supergravity agree precisely with those
computed via string theory provided that the matrix ${\cal M}$ is such
that
\be
{\cal M} = \frac{1}{\sqrt{2\D}} \left ( \begin{array} {c c }
(\D + 1)^{1/2} & - (\D - 1)^{1/2} \\
(\D - 1)^{1/2} & (\D + 1)^{1/2}
\end{array} \right )
\ee
for $\D = (n-1) \ge 2$. This agreement provides further evidence for
the non-renormalization theorem. Note however that the correspondence
between supergravity operators and those in the orbifold CFT is not
the naive relation one might have anticipated: ${\cal M}$ is not diagonal.
This explains the early observation that the cubic couplings in supergravity look very
different from the structure constants in the orbifold CFT three point
functions.

As discussed in \cite{Kanitscheider:2006zf} such a linear map between supergravity
and orbifold CFT operators is not sufficient to obtain matching for
all the extremal correlators. To understand how this issue is resolved,
one needs to recall the large $N$ scaling behavior of correlators: the key is
that extremal {\it non-linear} operator mixings are not suppressed in
extremal correlators \cite{D'Hoker:1999ea}.
Thus one can consider an identification between orbifold CFT
operators and supergravity operators of the form
\be
{\cal O}^{p,p}_{\Delta m \bar{m}} \leftrightarrow
{\cal O}^{\Phi}_{\Delta m \bar{m}} + \frac{1}{\sqrt{N}} \sum_{i, j}
b_{ij} {\cal O}^{\Phi_i}_{\Delta_i m_i \bar{m}_i} {\cal O}^{\Phi_j}_{ (\Delta
  - \Delta_i) (m-m_i) (\bar{m} - \bar{m}_i)} + \cdots,
\ee
where $b_{ij}$ are certain $N$ independent coefficients and the
ellipses denote subleading terms in $N$. Such a two particle term
contributes at leading order to certain extremal
three point functions, but only at subleading order to non-extremal three point
functions. Thus with suitable choices of $b_{ij}$ one can
match the extremal correlators computed in supergravity with those
computed in string theory and the orbifold theory.

The physical interpretation of such non-linear mixings is that single
particle string and orbifold CFT operators do not correspond to
single particle supergravity operators. At first sight this may seem
surprising, since one might have anticipated that the string
worldsheet vertex operators for supergravity modes would correspond to
single particle supergravity fields, as they do in flat space.
However, there is no contradiction: the matching between supergravity
fields and string vertex operators is defined by taking the
limit of the string computations of $n$-point functions,
and comparing with the corresponding supergravity computations. Thus
the comparison made here is the correct way to define the relationship
between operators dual to supergravity fields and string vertex operators.

\bigskip

So, to summarize, the non-extremal and extremal three point functions
computed in supergravity, in string theory and in the orbifold theory
agree provided that one correctly matches operators and takes into
account certain extremal non-linear operator mixings. Matching of the
correlators determines the map between supergravity and
orbifold/string theory operators, where quantum numbers alone do not
uniquely determine it.

One might wish to explore whether other correlators are protected by
non-renormalization theorems. In the analogous case of ${\cal N} =4$
SYM, there is evidence that extremal (and next to extremal)
$n$-point functions of chiral primaries are similarly protected, see
\cite{D'Hoker:2002aw}, and thus it is possible that all extremal
correlators in this case too will match between string theory, the
orbifold CFT and supergravity. Note however that general
non-extremal $n$-point functions for $n \ge 4$ are not protected
even in ${\cal N} = 4$ SYM, and are thus unlikely to be protected in
this less supersymmetric system. Comparison of the extremal
correlators will again be subtle since non-linear operator mixings
may contribute at leading order. That is, in an extremal $n$-point
function mixings of the type \be {\cal O}^{\Phi}_{\Delta} +
\frac{1}{N^{(n-2)/2}} \prod_{i=1}^{n-1} {\cal O}^{\Phi_i}_{\Delta_i}
+ \cdots \ee with $\sum_{i} \Delta_i = \Delta$ are not suppressed.

An important open issue is to understand better when there are
non-renormalization theorems for correlators. The (almost) proof of
the non-renormalization of three point functions in ${\cal N} = 4$
SYM relies on sophisticated harmonic superspace techniques. In this
case one should be able to use the $2d$ ${\cal N} = 4$ supersymmetry
to explain the non-renormalization. However, an
understanding of the non-renormalization from the bulk supergravity
perspective would more immediately generalize to other AdS/CFT
dualities. Such a non-renormalization theorem in the bulk would
involve arguing that $\a'$ corrections to the onshell renormalized
supergravity action do not contribute to the correlators.

The above discussion relates to three point functions of chiral
primaries associated with single particle supergravity fields. Given
that these appear to satisfy a non-renormalization theorem, with
appropriate operator mixing taken into account, it seems very likely
that all three point functions of multi-particle chiral primaries
are similarly protected. In the analogous case of ${\cal N} = 4$
SYM, there is indeed evidence for this, from both harmonic
superspace considerations \cite{Heslop:2001gp} and more recently
from the holographic analysis of LLM bubbling solutions in \cite{Skenderis:2007yb}.

If indeed all three point functions are protected, then an immediate
consequence would be that vevs of chiral primary operators in states
created by other chiral primaries are also not renormalized. Now in
\cite{Skenderis:2006ah, Kanitscheider:2006zf,Kanitscheider:2007wq}
such vevs were used to test the proposed correspondence between 1/2
BPS D1-D5 fuzzball geometries and superpositions of RR ground
states. Non-renormalization of the vevs can be used to push this
correspondence much further, as will be explored in a separate publication.
%

\bigskip

The plan of this paper is as follows. In section \ref{sugra1} we
review the results of the supergravity computation of three point
functions, whilst in section \ref{string} the results of the
corresponding string theory computations are reviewed. In section
\ref{non-ext} the non-extremal correlators are found to match, with an
appropriate linear identification of operators. In section \ref{ext}
it is shown that additional non-linear terms in this operator identification
are needed to obtain matching of extremal correlators.

\section{Supergravity computation of correlators} \la{sugra1}

In this section we will review the holographic computation of three point
functions of single particle scalar chiral primaries. These
correlators are computed by perturbing about the $AdS_3
\times S^3 \times X_4$ background, where $X_4$ is $T^4$ or
$K3$. There are three distinct families of scalar chiral primaries associated
with the $(p,p)$ cohomology of $X_4$ with $p=0,1,2$ respectively. The
operators couple to the following scalar fields in $AdS_3$:
\be
S_{kI}; \qquad S^{(r)}_{kI}; \qquad \Sigma_{kI}.
\ee
Here $(k,I)$ denote the $SO(4)$ R symmetry labels: $k$ is the degree
of the associated $S^3$ scalar spherical harmonic and $I$ denotes the
remaining Dynkin labels. Expressing $SO(4) = SU(2)_L \times SU(2)_R$,
appropriate labels are the $J^3_{L/R}$ charges $(m, \bar{m})$.
The middle cohomology of $X_4$  is labeled by
$(r) = 1, \cdots h^{1,1}(X_4)$ where $h^{1,1}$ is four for $T^4$ and
twenty for $K3$; this label defines the field transformations under
the global $SO(h^{1,1})$ symmetry.

Up to the overall normalization factor, the kinetic terms for these
fields are canonically normalized, namely the bulk action is
\bea \la{pp1}
{\cal S} &=& \frac{N}{4 \pi} \int d^{3}x \sqrt{-G} ( R_G + 2 -
\half ( (DS_{kI})^2 - k (k-2) (S_{kI})^2) \\
&& \qquad - \half ( (DS^{(r)}_{kI})^2 - k (k-2) (S^{(r)}_{kI})^2 )
- \half ( (D\S_{kI})^2 - k (k-2) \S_{kI}^2) + \cdots ). \nn
\eea
The overall normalization is proportional to the integer $N = n_1 n_5$.
Note that the mass terms are such that the scalar fields associated
with degree $k$ harmonics couple to operators of dimension $k$. For
the $S$ fields $k \ge 1$ whilst for the $\S$ fields $k \ge 2$.
To calculate the three point functions one also needs the appropriate cubic
couplings computed in \cite{Mihailescu:1999cj,Arutyunov:2000by}.
These are given by
\bea
&& - \frac{N}{4 \pi} \int d^3x \sqrt{-G} (T_{123}
S^{(a) 1} S^{(a)2} \S^3 + U_{123} \S^1 \S^2 \S^3); \la{3pcou} \\
&& \equiv - \frac{N}{16 \pi} \int d^3x \sqrt{-G} V_{123}
\left ( \frac{S^{(a)1} S^{(a)2} \S^3}{ \sqrt{(k_1
    +1)(k_2 + 1)}} + \frac{(k_1^2 + k_2^2 + k_3^2 -2)}{(k_1 + 1) (k_2+1)}
\frac{\S^1 \S^2 \S^3}{6 \sqrt{(k_1 -1)(k_2 -
    1)}} \right ), \nn \\
&& \hsp V_{123} = \frac{\S (\S+2) (\S-2)
  \a_1 \a_2 \a_3 a_{123}}{(k_3 + 1) \sqrt{k_1 k_2 k_3 (k_3 - 1)}} \nn
\eea
where $k_i$ denotes the dimension of the operator dual to the field $\Psi^i$,
$\S = k_1 + k_2 + k_3$, $\a_1 = \half (k_2 + k_3 - k_1)$ etc and $a_{123}$
is shorthand for the spherical harmonic overlap. Here the label $(a) =
1, \cdots h^{1,1}(X_4) + 1 \equiv n$ includes all $S$ fields. Compactification
of type IIB on $X_4$ gives rise to a theory with $SO(n)$ symmetry, and
the cubic couplings respect this symmetry. Note however that is an
accidental symmetry: only the $SO(h^{(1,1)})$ symmetry is respected by
the orbifold CFT and string theory three point functions.

The (renormalized) correlators can be computed using standard holographic
renormalization techniques. The two point functions are \cite{Kanitscheider:2006zf}:
\bea \la{two}
\langle {\cal O}^{S^{(a)}}_{k_1 I_1} (x) {\cal O}^{S^{(b)}}_{k_2 I_2} (0)
\rangle_h &=& \frac{N}{2 \pi^2} (k_1 - 1)^2 \left ( \frac{1}{x^{2k_1}}
  \right )_{R} \d_{I_1 I_2} \d_{k_1 k_2} \d^{(a) (b)}; \qquad k \neq 1
  \\
\langle {\cal O}^{\S}_{k_1 I_1} (x) {\cal O}^{\S}_{k_2 I_2} (0)
\rangle_h &=& \frac{N}{2 \pi^2} (k_1 - 1)^2 \left ( \frac{1}{x^{2k_1}}
  \right )_{R} \d_{I_1 I_2} \d_{k_1 k_2}, \nn
\eea
where ${\cal O}^{S^{(a)}}$ and ${\cal O}^{\S}$ denote the operators dual
to $S^{(a)}$ and $\S$ respectively. The subscript $R$ indicates that
the expressions are renormalized whilst the subscript $h$ in these and
subsequent expressions denotes that these are the holographically computed correlators.
When $k=1$, $(k-1) \rightarrow 1$ in the first
expression; recall that there is no $k=1$ operator ${\cal O}_{\S}$.

The three point functions are \cite{Kanitscheider:2006zf}:
\bea \la{three}
\< \cao^{S^{(a)}}(x_1) \cao^{S^{(b)}}(x_2) \cao^{\S}(x_3)\>_h &=& \frac{N}{4
  \pi^3} \frac{W_{123} T_{123} \d^{(a) (b)} }{|\vec{x}_1 - \vec{x}_2|^{2\a_3}
|\vec{x}_1 - \vec{x}_3|^{2 \a_2}
|\vec{x}_2 - \vec{x}_3|^{2 \a_1}}; \\
\< \cao^{\S}(x_1) \cao^{\S} (x_2) \cao^{\S} (x_3)\>_h &=& \frac{3 N}{4
  \pi^3} \frac{W_{123} U_{123}}{|\vec{x}_1 - \vec{x}_2|^{2\a_3}
|\vec{x}_1 - \vec{x}_3|^{2 \a_2}
|\vec{x}_2 - \vec{x}_3|^{2 \a_1}}; \nn \\
W_{123} &=& \frac{\G(\a_1) \G(\a_2)\G(\a_3 )\G(\half (\S-2))}{
\G(k_1 - 1)\G(k_2 -1)\G(k_3 -1)}. \nn
\eea
Here the operator at position $x_i$ has dimension $k_i$ and $SO(4)$ R
symmetry labels $I_i$.

To compare with the orbifold CFT and string theory computations one
wants normalized three point functions, dividing out by the
norms of the operators as given by the two point
functions. Suppressing the standard position dependence, this gives:
\bea
\< \hat{\cao}^{S^{(a)}} \hat{\cao}^{S^{(b)}}  \hat{\cao}^{\S} \>_h &=& \frac{1}{\sqrt{2N}}
\td{W}_{123} T_{123} \d^{(a) (b)}; \la{sugra} \\
\< \hat{\cao}^{\S} \hat{\cao}^{\S}  \hat{\cao}^{\S} \>_h &=& \frac{3}{\sqrt{2N}}
\td{W}_{123} U_{123}; \nn \\
\td{W}_{123} &=&  \frac{\G(\a_1) \G(\a_2)\G(\a_3 )\G(\half (\S-2))}{
\G(k_1)\G(k_2)\G(k_3)}, \nn
\eea
where $\hat{O}$ denotes the unit normalized operators. The remaining correlators
vanish
\be
\< \hat{\cao}^{S^{(a)}} \hat{\cao}^{S^{(b)}}  \hat{\cao}^{S^{(c)}} \>_h =
\< \hat{\cao}^{S^{(a)}} \hat{\cao}^{\S}  \hat{\cao}^{\S} \>_h = 0, \la{sugra2}
\ee
regardless of the operator dimension.

For later purposes it will be useful to give explicitly extremal correlators
in which one of the $\a_i = 0$. These are {\it defined} as the
continuation of the expressions (\ref{sugra}): the pole in
$\td{W}_{123}$ as one of the $\a_i \rightarrow 0$ cancels a
corresponding zero in the bulk couplings $(T_{123},U_{123})$ to give a
finite limit.
The relevant normalized extremal three point functions are thus of three types:
\bea \la{three-norm}
\< \hat{\cao}^{S^{(a)} \dagger}_{k_1 + k_2} \hat{\cao}^{S^{(b)}}_{k_1}
\hat{\cao}^{\S}_{k_2} \>_h &=& \d^{(a) (b)}
\frac{a_{123}}{\sqrt{N}}  \sqrt{\frac{2 k_1 k_2 (k_1 + k_2) (k_1 + k_2  + 1)}{ (k_1 + 1) (k_2 +
    1)^2  (k_2  -1)}}; \\
\< \hat{\cao}^{\S \dagger }_{k_1 + k_2}
\hat{\cao}^{S^{(a)}}_{k_1} \hat{\cao}^{S^{(b)}}_{k_2} \>_h &=&  \d^{(a) (b)}
\frac{a_{123}}{\sqrt{N}} \sqrt{\frac{2 k_1 k_2 (k_1 + k_2)}{(k_1 + 1) (k_2 +
    1)  (k_1 + k_2  -1)}}; \nn \\
\< \hat{\cao}^{\S \dagger}_{k_1 + k_2} \hat{\cao}^{\S}_{k_1} \hat{\cao}^{\S}_{k_2}
\>_h &=& \frac{a_{123}}{\sqrt{N}} \frac{(k_1^2 + k_2^2 + (k_1 +
k_2)^2 -2)^2}{(k_1 + 1) (k_2 + 1)}
\sqrt{\frac{k_1 k_2 (k_1 + k_2)} {2 (k_1  -1) (k_2  -1)(k_1 +k_2  -1) }}.  \nn
\eea
Note that the triple overlap $a_{123} = 1$ when the operator
with maximum dimension also has $SO(2)$ R charges which are minus the
sums of the $SO(2)$ charges of the other operators.
In particular that the extremal correlators at lowest dimension
are:
\be
\< \hat{\cao}^{\S \dagger }_{2}
\hat{\cao}^{S^{(a)}}_{1} \hat{\cao}^{S^{(b)}}_{1} \>_h =  \d^{(a) (b)}
\frac{a_{123}}{\sqrt{N}}; \qquad
\< \hat{\cao}^{S^{(c)} \dagger }_{2}
\hat{\cao}^{S^{(a)}}_{1} \hat{\cao}^{S^{(b)}}_{1} \>_h = 0.
\ee

\section{String theory/orbifold CFT correlators} \la{string}

In this section we will review the results for the corresponding
correlators computed in string theory and the orbifold CFT. Not all
non-extremal correlators have been computed in the orbifold CFT, but those
which are known agree with those computed via the string theory, as do
all extremal correlators. Here we summarize the results of
\cite{Gaberdiel:2007vu,Dabholkar:2007ey,Pakman:2007hn}
for the string theory computation of three point functions of
scalar chiral primaries. Note that general three point functions
involving vector chiral primaries associated with the $(0,2)$ and $(2,0)$
cohomology of $X_4$ are also given in \cite{Dabholkar:2007ey,Pakman:2007hn}.
We will not consider these here, since the corresponding holographic
correlators have not been computed, but it should be straightforward
to extend our discussions to these operators.

The scalar chiral primaries are labeled by the $(p,p)$ cohomology of
$X_4$, their twist $n \ge 1$, and their R symmetry $SU(2)_L \times
SU(2)_R$ quantum numbers as
\be
{\cal O}^{0,0}_{n m \bar{m}}; \qquad
{\cal O}^{(r)1,1}_{n m \bar{m}}; \qquad
{\cal O}^{2,2}_{n m \bar{m}}.
\ee
Here $(m,\bar{m})$ are the eigenvalues of $J^3$ and $\bar{J}^3$
respectively and $(r)$ labels the $(1,1)$ cohomology of $X_4$, of
dimension $h^{1,1}$; thus $(r)$
runs from $1$ to $4$ for $T^4$ and from $1$ to $20$ for $K3$. The
cohomology label is equivalent to giving the transformation properties
under the $SO(h^{1,1})$ global symmetry. The operator dimension is given by
\be
\Delta = (n - 1 + p),
\ee
with $J = \bar{J} = \half \Delta$. These operators are orthonormal
\be
\langle {\cal O}^{p,p}_{n -m -\bar{m}} {\cal O}^{p,p}_{n m \bar{m}}
\rangle_s = 1,
\ee
with the subscript $s$ denoting that these are string theory
correlators. The three point
functions can be conveniently expressed as
\bea
\< {\cal O}^{\ep_1,\bar{\ep}_1}_{n_1 m_1 \bar{m}_1}
{\cal O}^{\ep_2,\bar{\ep}_2}_{n_2 m_2 \bar{m}_2}
{\cal O}^{\ep_3,\bar{\ep}_3}_{n_3 m_3 \bar{m}_3} \>_s &=&
\frac{1}{\sqrt{N}} L(J_i,m_i) L(\bar{J}_i,\bar{m}_i)
\frac{ (\sum_{i=1}^3 \ep_i n_i + 1) (\sum_{i=1}^3 \bar{\ep}_i n_i +
  1)}{4 (n_1 n_2 n_3)^{1/2}}; \nn \\
\< {\cal O}^{(r) 1, 1}_{n_1 m_1 \bar{m}_1}
{\cal O}^{(s) 1,1}_{n_2 m_2 \bar{m}_2}
{\cal O}^{\ep,\bar{\ep}}_{n_3 m_3 \bar{m}_3} \>_s &=&
\frac{1}{\sqrt{N}} \d^{(r)(s)} L(J_i,m_i) L(\bar{J}_i,,\bar{m}_i)
\left ( \frac{ n_1 n_2}{n_3} \right )^{1/2}; \la{cft}
\eea
where $\ep = (p-1)$ for $p = 0,2$. Here
\bea
L(J_i,m_i) &=& d^{J_1,J_2,J_3}_{m_1,m_2,m_3} \eta_{J_i}
\left (  \frac{ \a_1 ! \a_2
  ! \a_3 ! (J_1 + J_2 + J_3 + 1)!}{(2 J_1)! (2J_2)! (2J_3)!} \right
)^{1/2}, \\
\eta_{J_i} &=& (-)^{\half (J_1 + J_2 + J_3)}. \nn
\eea
with $\a_1 = J_2 + J_3 - J_1 = \half (\Delta_2 + \Delta_3 - \Delta_1)$
etc and
\be
d^{J_1,J_2,J_3}_{m_1,m_2,m_3} = \left ( \begin{array}{ccc}  J_1 & J_2 & J_3 \\
  m_1 & m_2 & m_3   \end{array} \right ),
\ee
are the $SU(2)$ $3j$ symbols. Note that $U(1)$ R-charge conservation
enforces that $m_1 + m_2 + m_3 = \bar{m}_1 + \bar{m}_2 + \bar{m}_3 =
0$ in the correlators.

\section{Matching non-extremal correlators} \la{non-ext}

Let us now consider the matching of the correlators (\ref{sugra}),
(\ref{sugra2}) and (\ref{cft}). The most general {\it linear} identification
of operators which respects the symmetries is
\bea
{\Bbb O}^{S^{(r)}}_{n m  \bar{m}} & \leftrightarrow &
{\cal O}^{(r)1,1}_{n m \bar{m}};
\nn \\
\left ( \begin{array} {c} {\Bbb O}^{S}_{(n-1) m   \bar{m}} \\
{\Bbb O}^{\S}_{(n-1) m   \bar{m}} \end{array} \right )
&=& {\cal M}
\left ( \begin{array} {c} {\cal O}^{0,0}_{n m \bar{m}} \\
{\cal O}^{2,2}_{(n-2) m \bar{m}} \end{array} \right ),
\eea
for an arbitrary $SO(2)$ matrix ${\cal M}$. Here we denote by ${\Bbb
  O}^{\Phi}_{\D}$ operators which for {\it non-extremal} correlators
are to be identified with the holographic operators $\hat{\cal
  O}^{\Phi}_{\D}$. For {\it extremal} correlators we will need to
refine the map between these operators and the holographic operators.

The matrix ${\cal M}$ is completely fixed by the vanishing of the correlators
\be
\< {\Bbb O}^{S} {\Bbb O}^{S^{(r)}} {\Bbb O}^{S^{(s)}} \> = 0,
\ee
which implies
\be
{\cal M} = \frac{1}{\sqrt{2\D}} \left ( \begin{array} {c c }
(\D + 1)^{1/2} & - (\D - 1)^{1/2} \\
(\D - 1)^{1/2} & (\D + 1)^{1/2}
\end{array} \right )
\ee
for $\D = (n-1) \ge 2$. Clearly there can be no operator mixing at $\D =
1$, since there are no dimension one ${\Bbb O}^{\S}$ operators;
this is hence a special case which will be discussed separately.

Having determined ${\cal M}$, there is no further freedom in the
operator identification and one can check whether the remaining
correlators agree. Forming the appropriate linear combinations of the
string theory correlators (\ref{cft}), one finds that
\bea
\< {\Bbb O}^{S^{(a)}} {\Bbb O}^{S^{(b)}}  {\Bbb O}^{S^{(c)}} \>_{s}
&=& 0; \qquad \D_{i} \neq 1, \\
\< {\Bbb O}^{S^{(a)}} {\Bbb O}^{\S}  {\Bbb O}^{\S} \>_{s}
&=& 0; \qquad \D_{1} \neq 1, \nn \\
\< {\Bbb O}^{\S} {\Bbb O}^{S^{(a)}}  {\Bbb O}^{S^{(b)}} \>_{s}
&=& \frac{1}{\sqrt{N}} L(J_i,m_i) L(\bar{J}_i,\bar{m}_i) \d^{(a)(b)}
\frac{\sqrt{2} (\Delta_1 \Delta_2 \Delta_3)^{1/2}}{(\D_1^2 -
  1)^{1/2}}; \la{p1} \\
\< {\Bbb O}^{\S} {\Bbb O}^{\S}  {\Bbb O}^{\S} \>_{s} &=& \frac{(\D_1^2
  + \D_2^2 + \D_3^2 -2)}{2 ((\D_2^2 - 1) (\D_3^2 - 1))^{1/2}}
\< {\Bbb O}^{\S} {\Bbb O}^{S^{(a)}}  {\Bbb O}^{S^{(b)}} \>_{s}. \la{p2}
\eea
Here the subscript $s$ denotes that these are linear combinations of
the correlators computed in the string theory.

Now let us compare these correlators with the holographic
correlators (\ref{sugra}) and
(\ref{sugra2}): the zeroes given in (\ref{sugra2}) are reproduced
(except in the special cases involving dimension one
operators). Moreover, using (\ref{sugra}) and noting that
\be
\frac{3 U_{123}}{T_{231}} = \frac{ ( \D_1^2 + \D_2^2 + \D_3^2  - 2)}{2
  ( (\D_2^2 - 1)(\D_3^2 - 1))^{1/2}},
\ee
one sees that the ratio given in (\ref{p2}) indeed agrees with that
from supergravity. Thus one need only compare the overall
normalization of (\ref{p1}) with that of the supergravity correlator
for all non-extremal correlators to match. Noting that
\be
\td{W}_{123} T_{231} = \frac{ \a_1 ! \a_2! \a_3! (J_1 + J_2 + J_3 +
  1)!}{(2 J_1)! (2 J_2)! (2 J_3)!} \frac{ \sqrt{\D_1 \D_2 \D_3}}{
  (\D_1^2 - 1)^{1/2}} \frac{2 a_{123}}  {( (\D_1 + 1 ) (\D_2 + 1)
  (\D_3 + 1)  )^{1/2}}, \nn
\ee
the holographic correlator (\ref{sugra}) can be rewritten in terms of
the string theory correlator as
\be
\frac{\< \hat{\cal O}^{\S} \hat{\cal O}^{S^{(a)}}  \hat{\cal O}^{S^{(b)}}
  \>_h}{\< {\Bbb O}^{\S} {\Bbb O}^{S^{(a)}}  {\Bbb O}^{S^{(b)}}
  \>_s}
= \frac{1}{\eta_{J_i} \eta_{\bar{J}_i}
d^{J_1,J_2 J_3}_{m_1,m_2,m_3}  d^{\bar{J}_1,\bar{J}_2
  \bar{J}_3}_{\bar{m}_1,\bar{m}_2, \bar{m}_3}}
\frac{a_{123}} {{( (\D_1 + 1 ) (\D_2 + 1) (\D_3 + 1)  )^{1/2}}}.
\ee
Triple integrals of spherical harmonics can be expressed in terms of
$3j$ symbols; in particular the triple overlap $a_{123}$ can be
written as \cite{Cutkosky:1983jd}
\be
a_{123} = \eta_{J_i} \eta_{\bar{J}_i}
d^{J_1,J_2 J_3}_{m_1,m_2,m_3}  d^{\bar{J}_1,\bar{J}_2
\bar{J}_3}_{\bar{m}_1,\bar{m}_2, \bar{m}_3}
{( (2J_1 + 1 ) (2 J_2 + 1) (2 J_3 + 1)  )^{1/2}},
\ee
and thus the normalization of the holographic correlators precisely
matches that of the string correlators!

\section{Matching of exceptional extremal correlators} \la{ext}

The linear matching between supergravity and orbifold CFT operators is
sufficient for all non-extremal correlators, and most extremal
correlators, to match. There remains however a discrepancy for
correlators involving dimension one operators, where no linear mixing
was possible.

For correlators which involve at least one dimension one operator,
some agree with the holographic results, namely
\bea
\< {\Bbb O}^{S}_{\D_2} {\Bbb O}^{S^{(r)}}_1
{\Bbb O}^{S^{(s)}}_{\D_3} \>_{s}
&=& 0; \\
\< {\Bbb O}^{\S}_{\D_2} {\Bbb O}^{S^{(r)}}_1
{\Bbb O}^{S^{(s)}}_{\D_3} \>_{s}
&=& \frac{1}{\sqrt{N}} L(J_i,m_i) L(\bar{J}_i,\bar{m}_i) \d^{(r) (s)}
  \frac{ \sqrt{2 \D_2 \D_3}}{(\D_2^2 -1)^{1/2}}; \nn \\
\< {\Bbb O}^{S}_{1} {\Bbb O}^{S}_{\D_2}
{\Bbb O}^{\S}_{\D_3} \>_{s} &=& \frac{1}{\sqrt{N}} L(J_i,m_i) L(\bar{J}_i,\bar{m}_i)
\frac{ \sqrt{2 \D_2 \D_3}}{(\D_3^2 -1)^{1/2}}, \nn
\eea
but the rest do not:
\bea
\< {\Bbb O}^{S}_{1} {\Bbb O}^{S}_{1}
{\Bbb O}^{S}_{2} \>_{s} &=& \frac{2}{\sqrt{N}} L(J_i,m_i)
L(\bar{J}_i,\bar{m}_i); \qquad
\< {\Bbb O}^{S}_{1} {\Bbb O}^{S}_{1}
{\Bbb O}^{\S}_{2} \>_{s} = \frac{\sqrt{3}}{\sqrt{N}} L(J_i,m_i)
L(\bar{J}_i,\bar{m}_i); \nn \\
\< {\Bbb O}^{S}_{1} {\Bbb O}^{S^{(r)}}_{\D_2}
{\Bbb O}^{S^{(s)}}_{\D_3} \>_{s}
&=& \frac{1}{\sqrt{N}} L(J_i,m_i) L(\bar{J}_i,\bar{m}_i) \left ( \frac{\D_2
  \D_3}{2} \right )^{1/2}; \la{tr1} \\
\< {\Bbb O}^{S}_{1} {\Bbb O}^{S}_{\D_2}
{\Bbb O}^{S}_{\D_3} \>_s &=& \frac{1}{\sqrt{2N}}
L(J_i,m_i) L(\bar{J}_i,\bar{m}_i) \sqrt{ \D_2 \D_3}; \nn \\
\< {\Bbb O}^{S}_{1} {\Bbb O}^{\S}_{\D_2}
{\Bbb O}^{\S}_{\D_3} \>_s &=& \frac{1}{2 \sqrt{2N}}
L(J_i,m_i) L(\bar{J}_i,\bar{m}_i) \frac{\sqrt{ \D_2 \D_3}}{ ((\D_2^2 -
  1)(\D_3^2 - 1))^{1/2}} (\D_2^2 + \D_3^2 -1), \nn
\eea
with the corresponding holographic correlators being
\bea
\< \hat{\cao}^{S}_{1} \hat{\cao}^{S}_{\D_2}
\hat{\cao}^{S}_{\D_3} \>_{h} &=& \< \hat{\cao}^{S}_{1} \hat{\cao}^{S^{(r)}}_{\D_2}
\hat{\cao}^{S^{(s)}}_{\D_3} \>_{h} =
\< \hat{\cao}^{S}_{1} \hat{\cao}^{\S}_{\D_2}
\hat{\cao}^{\S}_{\D_3} \>_h = 0; \\
\< \hat{\cao}^{S}_{1} \hat{\cao}^{S}_{1}
\hat{\cao}^{\S}_{2} \>_{h} &=& \frac{1}{\sqrt{N}} a_{123} \equiv
\frac{2}{\sqrt{3N}} L(J_i,m_i)
L(\bar{J}_i,\bar{m}_i). \nn
\eea
Note that all these correlators are extremal because the spherical harmonic
triple overlaps are only non-zero when $\D_3 = (\D_2 \pm 1)$. This
follows from the addition of $SO(4)$ representations
\be
\left (\frac{1}{2}, \frac{1}{2} \right ) \oplus \left (\frac{\D_2}{2},
\frac{\D_2}{2} \right )
\rightarrow \left (\frac{\D_2 \pm 1}{2}, \frac{\D_2 \pm 1}{2} \right ).
\ee
Whilst the extremal holographic correlators are (by construction) the analytic
continuation of corresponding non-extremal correlators, the string
theory correlators in (\ref{tr1}) are not the analytic continuation of
corresponding non-extremal correlators given in (\ref{p1}).
As we will now explain, this apparent discrepancy between extremal correlators
can be resolved by allowing for {\it non-linear} operator mixing.

\subsection{Large $N$ behavior of correlators}

Let ${\cal O}^{\Phi}_{k}$ denote the operator of dimension $k$
and $SO(2)$ R charges $(k/2,k/2)$ dual to the supergravity
field $\Phi$, where $\Phi = (S^{(a)},\S)$. Now denote by
\be
{\cal O}^{[\Phi]_n}_k
= \left [ \prod_{i=1}^n {\cal O}^{\Phi_i}_{k_i} \right ]
\ee
the associated protected $n$-particle operators, with dimension
$k = \sum_i k_i$ and R charges $(k/2, k/2)$. Here $[{\cal
    O}^{\Phi_i}_{k_i} \cdots ]$ denotes the highest weight component
of the direct product of $SO(4)$ representations.

The operators ${\cal O}^{[\Phi]_n}_k$ transform in the same $SO(4)$
representation as the single particle operators ${\cal O}^{\Phi}_{k}$
and therefore one would anticipate that there is operator
mixing. Although generically operator mixing with multi-particle
operators is suppressed in the large $N$ limit, this is not true for
operators transforming in the same representations. One can understand
this from large $N$ counting arguments as follows.

Consider first correlation functions of single particle operators.
The two and three point functions computed from gravity scale as $N$,
as given in (\ref{two}) and (\ref{three}),
so the normalized two and three point functions
scale as one and $1/\sqrt{N}$ respectively.  Four point functions
include both disconnected and connected contributions. The former
scale as $N^2$ and are such that
\be
\< {\cal O}^{\Phi_1}_{k_1}  (x_1) {\cal O}^{\Phi_2}_{k_2} (x_2) {\cal
  O}^{\Phi_3}_{k_3} (x_3) {\cal O}^{\Phi_4}_{k_4} (x_4) \> = N^2 \left
( \d^{\Phi_1 \Phi_2} \d^{\Phi_3 \Phi_4} \frac{\d(k_1 + k_2) \d(k_3 +
  k_4)}{x_{12}^{2|k_1|} x_{34}^{2|k_3|}} + \cdots \right ),
\ee
where the ellipses denote permutations and numerical factors are
suppressed. The scaling as $N^2$ follows from the fact that these
disconnected contributions are the products of two point functions.
Working with unit normalized operators, the disconnected contribution
to the four point function thus scales as one. Connected contributions
to four point functions however scale as $N$ or, working with unit
normalized operators, as $1/N$. Note that holographic computation of the connected
contributions involves both the cubic and quartic couplings, whilst
the disconnected contributions follow entirely from the (renormalized)
quadratic action.

Now let us consider correlation functions involving multi particle
operators. In particular, one can read off the large $N$ behavior of
correlators involving double particle operators from the single
particle correlators discussed above. The operator product expansion
${\cal O}^{\Phi_1}_{k_1} (x_1) {\cal O}^{\Phi_2}_{k_2} (x_2) $
contains the term
\be
{\cal O}^{\Phi_1}_{k_1} (x_1) {\cal O}^{\Phi_2}_{k_2} (x_2)
\rightarrow [{\cal O}^{\Phi_1}_{k_1} (x_1) {\cal O}^{\Phi_2}_{k_2} (x_1)]
\ee
with unit coefficient as $x_1 \rightarrow x_2$ since the double
particle operator is defined by the short distance limit. Thus from
the $x_1 \rightarrow x_2$ behavior of correlators one can extract the
$N$ scaling of mixed correlators involving both single and multi
particle operators. Working with unit normalized single particle
operators this gives
\bea
\< \hat{\cal O}^{\dagger \Phi_1}_{k}  (x_1) [\hat{\cal O}^{\Phi_2}_{k_1}
    (x_2) \hat{\cal O}^{\Phi_3}_{k_2} (x_2)] \> & \approx & \frac{C^{123}_{\; k_1 k_2}}{\sqrt{N}}; \\
\< [\hat{\cal O}^{\Phi_1}_{l} (x_1) \hat{\cal O}^{\Phi_2}_{k - l} (x_1)]^{\dagger}
\hat{\cal O}^{\Phi_3}_{k_1} (x_2) \hat{\cal O}^{\Phi_4}_{k_2} (x_3) \>
& \approx & ( \d^{l k_1} \d^{\Phi_1 \Phi_3} \d^{\Phi_2 \Phi_4}
+ \d^{l k_2} \d^{\Phi_1 \Phi_4} \d^{\Phi_2 \Phi_3}); \nn \\
\< [\hat{\cal O}^{\Phi_1}_{l} (x_1) \hat{\cal O}^{\Phi_2}_{k - l} (x_1)]^{\dagger}
\hat{\cal O}^{\Phi_3}_{k_1} (x_2) \hat{\cal O}^{\Phi_4}_{k_2} (x_3) \>
& \approx & \frac{1}{N}; \qquad l \neq k_1,k_2; \nn \\
\< [\hat{\cal O}^{\Phi_1}_{l} (x_1) \hat{\cal O}^{\Phi_2}_{k - l} (x_1)]^{\dagger}
[\hat{\cal O}^{\Phi_3}_{k_1} (x_2) \hat{\cal O}^{\Phi_4}_{k_2} (x_2)] \>
& \approx & ( \d^{l k_1} \d^{\Phi_1 \Phi_3} \d^{\Phi_2 \Phi_4}
+ \d^{l k_2} \d^{\Phi_1 \Phi_4} \d^{\Phi_2 \Phi_3}); \nn \\
\< [\hat{\cal O}^{\Phi_1}_{l} (x_1) \hat{\cal O}^{\Phi_2}_{k - l}
(x_1)]^{\dagger} [\hat{\cal O}^{\Phi_3}_{k_1} (x_2) \hat{\cal
O}^{\Phi_4}_{k_2} (x_2)] \> & \approx & \frac{1}{N}; \qquad l \neq
k_1,k_2. \nn
\eea
where in all cases $k = k_1 + k_2$, and thus the
correlators are extremal. The (standard) $x_i$ dependence of the
correlators is suppressed. Here structure constants $C^{123}_{k_1
k_2}$ follow from the extremal single particle correlators given in
(\ref{three-norm}). The second and fourth correlators follow from
the disconnected components of the four point functions, whilst the
third and fifth correlators pick up contributions only from the
connected components and are thus subleading.


This large $N$ counting demonstrates that
operator mixings which are extremal are not suppressed in extremal
correlators. That is, suppose one considers operators such that
\be
(\td{\cal O}^{\Phi_a}_{k_b+k_c}) = \hat{\cal O}^{\Phi_a}_{k_b+k_c} +
\frac{1}{\sqrt{N}} b^{abc}_{k_b k_c} [\hat{\cal O}^{\Phi_b}_{k_b} \hat{\cal
    O}^{\Phi_c}_{k_c} ]
+ \cdots, \la{expa}
\ee
where the ellipses denote three particle and higher mixings.
Then by construction
\bea
\< (\td{\cal O}^{\Phi_a}_{k_a})^{\dagger }
(\td{\cal O}^{\Phi_b}_{k_b}) \>  &=& \d^{\Phi_a \Phi_b} \d_{k_a k_b} + {\cal
  O}(\frac{1}{N}); \\
\< (\td{\cal O}^{\Phi_a}_{k_b+k_c})^{\dagger} (\td{\cal O}^{\Phi_b}_{k_b})
(\td{\cal O}^{\Phi_c}_{k_c}) \> &=& \frac{1}{\sqrt{N}} (
C^{a b c}_{k_b k_c} +  b^{abc}_{k_b k_c} ) + {\cal
  O}(\frac{1}{N}); \nn \\
&\equiv & \frac{1}{\sqrt{N}}  \td{C}^{ a b c}_
{k_b k_c} +  {\cal O}(\frac{1}{N}). \nn
\eea
Thus to leading order in $N$ the mixed operators have the same two
point functions as the single particle operators, and
their three point functions still scale as $1/\sqrt{N}$. However, the
structure constants are modified: $ C^{a b c}_{k_b k_c}
\rightarrow \td{C}^{a b c}_{k_b k_c}$. Note that the $N$ scaling of the $m$-particle
term in the mixing (\ref{expa}) is $1/N^{m/2}$ such that the $n$-point functions
of the mixed operators scale as $N^{(1 - n/2)}$.

In the case at hand, for the exceptional extremal
holographic and string correlators to agree, one needs the following
quadratic operator mixings:
\bea
{\Bbb O}^{\S}_{2} &=& \hat{\cal O}^{\S}_{2} + \frac{1}{ 2 \sqrt{3N}}
L(J_i,m_i) L(\bar{J}_i,\bar{m}_i) \hat{\cal O}^{S}_{1} \hat{\cal O}^{S}_{1} + \cdots;
\la{nlmix} \\
{\Bbb O}^{S}_{2}
&=& \hat{\cal O}^{S}_{2} + \frac{1}{\sqrt{N}}
L(J_i,m_i) L(\bar{J}_i,\bar{m}_i)
\hat{\cal O}^{S}_{1} \hat{\cal O}^{S}_{1} + \cdots;
\nn \\
{\Bbb O}^{S}_{\D+1} &=& \hat{\cal O}^{S}_{\D+1}
+ \frac{\sqrt{\D (\D+1)}} {\sqrt{2 N}}
L(J_i,m_i) L(\bar{J}_i,\bar{m}_i) \hat{\cal O}^{S}_{1} \hat{\cal O}^{S}_{\D} + \cdots;
\nn \\
{\Bbb O}^{S^{(r)}}_{\D+1}
&=& \hat{\cal O}^{S^{(r)}}_{\D+1}
+ \frac{\sqrt{\D (\D+1)}} {\sqrt{2 N}}
L(J_i,m_i) L(\bar{J}_i,\bar{m}_i) \hat{\cal O}^{S}_{1} \hat{\cal O}^{S^{(r)}}_{\D} + \cdots;
\nn \\
{\Bbb O}^{\S}_{\D+1} &=& \hat{\cal O}^{\S}_{\D+1} + \frac{1}{\sqrt{2 N}}
\frac{\D(\D+1)}{ \sqrt{(\D-1)(\D+2)}} L(J_i,m_i)
L(\bar{J}_i,\bar{m}_i) \hat{\cal O}^{S}_{1}
\hat{\cal O}^{\S}_{\D} + \cdots, \nn
\eea
where in the latter three cases $\D \ge 2$. The ellipses denote additional potential mixings, which
include both two particle operators involving vector chiral primaries and $n$-particle operators
with $n \ge 3$.

Several computations could in principle be used to verify the consistency of these
operator identifications. Firstly, one could compute finite $N$ corrections to
the supergravity and string theory/orbifold CFT two and three point
functions, although on the supergravity side this is currently
intractable since only a subset of the requisite corrections to the
effective action are known. Secondly, one could compute correlation
functions for operators in the same supermultiplets, which are dual to
other supergravity fields. These should also be protected, and the
operator identifications required for supergravity and string
theory/orbifold CFT correlations functions to agree
should descend from those given in (\ref{nlmix}).

As previously mentioned, it would be interesting to understand the
non-renormalization better, both from the perspective of the 2d ${\cal N}
= 4$ CFT and from supergravity. This could lead to other
non-renormalization theorems and give insights into the
required operator matching. More generally one would like to explore
further the relationship between
the supergravity and string theory computations, to
understand better the latter. In supergravity there is by now a
deep understanding of the holographic renormalization used to remove
infinite volume divergences and obtain renormalized
correlators. The same volume renormalization is also responsible for the
finiteness of correlators in the string computations, but
renormalization has not been systematically developed and applied
in this context. Moreover, in supergravity there is a natural
geometric understanding of
the connection between boundary conditions for bulk fields and dual
operators, whilst in the string computations the relation proposed in
\cite{Kut} between
worldsheet vertex operators and CFT operators is less well
understood. Thus insights from the supergravity holographic computations may
help to understand further the (successful) hypotheses used in the string
computations.

\section*{Acknowledgments}

The author would like to thank Ingmar Kanitscheider and Kostas
Skenderis for useful discusssions, and the Simons Workshop for
hospitality during the completion of this work. The author is supported
by NWO, via the Vidi grant ``Holography, duality and time dependence in string theory''.


\begin{thebibliography}{99}


\bibitem{Lee:1998bx}
  S.~Lee, S.~Minwalla, M.~Rangamani and N.~Seiberg,
  ``Three-point functions of chiral operators in D = 4, N = 4 SYM at  large
  N,''
  Adv.\ Theor.\ Math.\ Phys.\  {\bf 2} (1998) 697
  [arXiv:hep-th/9806074].

\bibitem{Heslop:2001gp} K.~A.~Intriligator,
  ``Bonus symmetries of N = 4 super-Yang-Mills correlation functions via  AdS
  duality,'' Nucl.\ Phys.\  B {\bf 551} (1999) 575
  [arXiv:hep-th/9811047];  K.~A.~Intriligator and W.~Skiba,
  ``Bonus symmetry and the operator product expansion of N = 4
  super-Yang-Mills,'' Nucl.\ Phys.\  B {\bf 559} (1999) 165
  [arXiv:hep-th/9905020];
   B.~Eden, P.~S.~Howe and P.~C.~West,
  ``Nilpotent invariants in N = 4 SYM,''
  Phys.\ Lett.\  B {\bf 463} (1999) 19
  [arXiv:hep-th/9905085];
A.~Petkou and K.~Skenderis,
  ``A non-renormalization theorem for conformal anomalies,''
  Nucl.\ Phys.\  B {\bf 561} (1999) 100
  [arXiv:hep-th/9906030];
P.~S.~Howe, C.~Schubert, E.~Sokatchev and P.~C.~West,
  ``Explicit construction of nilpotent covariants in N = 4 SYM,''
  Nucl.\ Phys.\  B {\bf 571} (2000) 71
  [arXiv:hep-th/9910011];
P.~J.~Heslop and P.~S.~Howe,
  ``OPEs and 3-point correlators of protected operators in N = 4 SYM,''
  Nucl.\ Phys.\  B {\bf 626}, 265 (2002)
  [arXiv:hep-th/0107212].

\bibitem{D'Hoker:1999ea}
  E.~D'Hoker, D.~Z.~Freedman, S.~D.~Mathur, A.~Matusis and L.~Rastelli,
  ``Extremal correlators in the AdS/CFT correspondence,''
  arXiv:hep-th/9908160.

\bibitem{D'Hoker:2002aw}
  E.~D'Hoker and D.~Z.~Freedman,
  ``Supersymmetric gauge theories and the AdS/CFT correspondence,''
  arXiv:hep-th/0201253.

\bibitem{David:2002wn}
  J.~R.~David, G.~Mandal and S.~R.~Wadia,
  ``Microscopic formulation of black holes in string theory,''
  Phys.\ Rept.\  {\bf 369}, 549 (2002)
  [arXiv:hep-th/0203048].


\bibitem{Dijkgraaf:1998gf}
  R.~Dijkgraaf,
  ``Instanton strings and hyperKaehler geometry,''
  Nucl.\ Phys.\  B {\bf 543}, 545 (1999)
  [arXiv:hep-th/9810210].

\bibitem{Larsen:1999uk}
  F.~Larsen and E.~J.~Martinec,
  ``U(1) charges and moduli in the D1-D5 system,''
  JHEP {\bf 9906}, 019 (1999)
  [arXiv:hep-th/9905064].

\bibitem{de Boer:1998ip}
  J.~de Boer,
  ``Six-dimensional supergravity on S**3 x AdS(3) and 2d conformal field
  theory,''
  Nucl.\ Phys.\  B {\bf 548} (1999) 139
  [arXiv:hep-th/9806104].

\bibitem{Deger:1998nm}
  S.~Deger, A.~Kaya, E.~Sezgin and P.~Sundell,
  ``Spectrum of D = 6, N = 4b supergravity on AdS(3) x S(3),''
  Nucl.\ Phys.\  B {\bf 536}, 110 (1998)
  [arXiv:hep-th/9804166].

\bibitem{Jevicki:1998bm}
  A.~Jevicki, M.~Mihailescu and S.~Ramgoolam,
  ``Gravity from CFT on S**N(X): Symmetries and interactions,''
  Nucl.\ Phys.\ B {\bf 577}, 47 (2000)
  [arXiv:hep-th/9907144].


\bibitem{Lunin:2001pw}
  O.~Lunin and S.~D.~Mathur,
  ``Three-point functions for M(N)/S(N) orbifolds with N = 4 supersymmetry,''
  Commun.\ Math.\ Phys.\  {\bf 227} (2002) 385
  [arXiv:hep-th/0103169]; O.~Lunin and S.~D.~Mathur,
  ``Correlation functions for M(N)/S(N) orbifolds,''
  Commun.\ Math.\ Phys.\  {\bf 219}, 399 (2001)
  [arXiv:hep-th/0006196].

\bibitem{Mihailescu:1999cj}
  M.~Mihailescu,
  ``Correlation functions for chiral primaries in D = 6 supergravity on  AdS(3)
  x S(3),''
  JHEP {\bf 0002}, 007 (2000)
  [arXiv:hep-th/9910111].

\bibitem{Arutyunov:2000by}
  G.~Arutyunov, A.~Pankiewicz and S.~Theisen,
  ``Cubic couplings in D = 6 N = 4b supergravity on AdS(3) x S(3),''
  Phys.\ Rev.\ D {\bf 63} (2001) 044024
  [arXiv:hep-th/0007061].

\bibitem{Pank}  A.~Pankiewicz, ``Six-dimensional supergravities and the
AdS/CFT correspondence,'' Diploma Thesis, University of Munich,
October 2000.


\bibitem{S2}
  S.~de Haro, S.~N.~Solodukhin and K.~Skenderis,
  ``Holographic reconstruction of spacetime and renormalization in the  AdS/CFT
  correspondence,''
  Commun.\ Math.\ Phys.\  {\bf 217}, 595 (2001)
  [arXiv:hep-th/0002230];
  M.~Bianchi, D.~Z.~Freedman and K.~Skenderis,
  ``How to go with an RG flow,''
  JHEP {\bf 0108}, 041 (2001)
  [arXiv:hep-th/0105276];
%
  M.~Bianchi, D.~Z.~Freedman and K.~Skenderis,
  ``Holographic renormalization,''
  Nucl.\ Phys.\ B {\bf 631}, 159 (2002)
  [arXiv:hep-th/0112119];
  D.~Z.~Freedman, S.~D.~Mathur, A.~Matusis and L.~Rastelli,
  ``Correlation functions in the CFT($d$)/AdS($d+1$) correspondence,''
  Nucl.\ Phys.\ B {\bf 546}, 96 (1999)
  [arXiv:hep-th/9804058];
  K.~Skenderis,
  ``Lecture notes on holographic renormalization,''
  Class.\ Quant.\ Grav.\  {\bf 19} (2002) 5849
  [arXiv:hep-th/0209067];
  I.~Papadimitriou and K.~Skenderis,
  ``AdS / CFT correspondence and geometry,''
  arXiv:hep-th/0404176;
  I.~Papadimitriou and K.~Skenderis,
  ``Correlation functions in holographic RG flows,''
  JHEP {\bf 0410}, 075 (2004)
  [arXiv:hep-th/0407071].

\bibitem{Skenderis:2006uy}
  K.~Skenderis and M.~Taylor,
  ``Kaluza-Klein holography,''
  JHEP {\bf 0605}, 057 (2006)
  [arXiv:hep-th/0603016].

\bibitem{Skenderis:2006ah}
  K.~Skenderis and M.~Taylor,
  ``Fuzzball solutions and D1-D5 microstates,''
  Phys.\ Rev.\ Lett.\  {\bf 98} (2007) 071601
  [arXiv:hep-th/0609154].

\bibitem{Kanitscheider:2006zf}
  I.~Kanitscheider, K.~Skenderis and M.~Taylor,
  ``Holographic anatomy of fuzzballs,''
  JHEP {\bf 0704} (2007) 023
  [arXiv:hep-th/0611171].

\bibitem{Kanitscheider:2007wq}
  I.~Kanitscheider, K.~Skenderis and M.~Taylor,
  ``Fuzzballs with internal excitations,''
  JHEP {\bf 0706}, 056 (2007)
  [arXiv:0704.0690 [hep-th]].

\bibitem{Skenderis:2007yb}
  K.~Skenderis and M.~Taylor,
  ``Anatomy of bubbling solutions,'' JHEP {\bf 0709} (2007) 019
  [arXiv:0706.0216 [hep-th]] .

\bibitem{Freedman:1998tz}
  D.~Z.~Freedman, S.~D.~Mathur, A.~Matusis and L.~Rastelli,
  ``Correlation functions in the CFT($d$)/AdS($d+1$) correspondence,''
  Nucl.\ Phys.\ B {\bf 546}, 96 (1999)
  [arXiv:hep-th/9804058].

\bibitem{Gaberdiel:2007vu}
  M.~R.~Gaberdiel and I.~Kirsch,
  ``Worldsheet correlators in AdS(3)/CFT(2),''
  JHEP {\bf 0704} (2007) 050
  [arXiv:hep-th/0703001].

\bibitem{Dabholkar:2007ey}
  A.~Dabholkar and A.~Pakman,
  ``Exact chiral ring of AdS(3)/CFT(2),''
  arXiv:hep-th/0703022.

\bibitem{Pakman:2007hn}
  A.~Pakman and A.~Sever,
  ``Exact N=4 correlators of AdS(3)/CFT(2),''
  arXiv:0704.3040 [hep-th].


\bibitem{Cutkosky:1983jd}
  R.~E.~Cutkosky,
  ``Harmonic Functions And Matrix Elements For Hyperspherical Quantum Field
  Models,''
  J.\ Math.\ Phys.\  {\bf 25} (1984) 939.

\bibitem{Kut}A.~Giveon, D.~Kutasov and N.~Seiberg,
  ``Comments on string theory on AdS(3),''
  Adv.\ Theor.\ Math.\ Phys.\  {\bf 2} (1998) 733
  [arXiv:hep-th/9806194]; D.~Kutasov, F.~Larsen and R.~G.~Leigh,
  ``String theory in magnetic monopole backgrounds,''
  Nucl.\ Phys.\  B {\bf 550} (1999) 183
  [arXiv:hep-th/9812027]; J.~de Boer, H.~Ooguri, H.~Robins and J.~Tannenhauser,
  ``String theory on AdS(3),''
  JHEP {\bf 9812} (1998) 026
  [arXiv:hep-th/9812046]; D.~Kutasov and N.~Seiberg,
  ``More comments on string theory on AdS(3),''
  JHEP {\bf 9904} (1999) 008
  [arXiv:hep-th/9903219]; J.~M.~Maldacena and H.~Ooguri,
  ``Strings in AdS(3) and the SL(2,R) WZW model. III: Correlation  functions,''
  Phys.\ Rev.\  D {\bf 65} (2002) 106006
  [arXiv:hep-th/0111180].











\end{thebibliography}
\end{document}